\documentclass[aps,twocolumn,superscriptaddress,pre,a4paper]{revtex4-1}

\fontencoding{T1}
\usepackage{epsfig}
\usepackage{amsfonts}
\usepackage{amsmath}
\usepackage{amssymb}
\usepackage{bm}
\usepackage{color}
\usepackage{graphicx}

\begin{document}

\title{Collective dynamics of pedestrians in a non-panic evacuation scenario}

\author{Juan Cruz {Moreno}}
\affiliation{Departamento de Ciencia y Tecnolog\'\i{}a, Universidad Nacional Quilmes, Argentina.}
\affiliation{Consejo Nacional de Investigaciones Cient\'\i{}ficas y Tecnol\'ogicas (CONICET), Argentina}
\affiliation{Ingeniería en Inform\'atica, Departamento de Tecnolog\'\i{}a y Administraci\'on,\\ Universidad Nacional de Avellaneda, Argentina}

\author{M.\ Leticia {Rubio Puzzo}}
\affiliation{Instituto de F\'\i{}sica de L\'\i{}quidos y Sistemas Biol\'ogicos (IFLYSIB), 
CONICET and Universidad Nacional de La Plata, Calle 59 no.~789, B1900BTE La Plata, Argentina}
\affiliation{CCT CONICET La Plata, Consejo Nacional de Investigaciones Cient\'\i{}ficas y T\'ecnicas, Argentina}
\affiliation{Departamento de F\'\i{}sica, Facultad de Ciencias Exactas,
  Universidad Nacional de La Plata, Argentina}

\author{Wolfgang {Paul}}
\affiliation{Institut f\"ur Physik, Martin-Luther-University Halle-Wittenberg,
06099 Halle, Germany}

\begin{abstract}

We present a study of pedestrian motion along a corridor in a non-panic regime,
as usually happens in evacuation scenarios in, e.g., schools, hospitals or
airports. Such situations have been discussed so far within the so-called
Social Force Model (SFM), a particle-based model with interactions depending
on the relative position of the particles. We suggest to enrich this model by
interactions based on the velocity of the particles and some randomness, both
of which we introduce using the ideas of the Vicsek Model (VM). This new model allows to
introduce fluctuations for a given average speed and geometry (because in 
real-life there are different evacuation modes at the same average speed), and
considering that the alignment interactions are modulated by an external
control parameter (the noise $\eta$) allows to introduce phase transitions between
ordered and disordered states. 

To clarify the influence of the model ingredients we have compared simulations
of pedestrian motion along a corridor using (a) the pure Vicsek Model (VM) with two boundary
conditions (periodic and bouncing back) and with or without desired direction
of motion, (b) the Social Force Model (SFM), and (c) the new model
developed as a combination of both (SFM+VM). 

The study of steady-state particle configurations in the VM with confined geometry
shows the expected bands perpendicular to the motion direction, while in the SFM
and SFM+VM particles order in stripes of a given width $w$ along the
direction of motion. The results in the SFM+VM case show that $w(t)\simeq t^\alpha$ has a
diffusive-like behavior at low noise $\eta$ (dynamic exponent $\alpha
\approx 1/2$), while it is sub-diffusive at high values of external noise
($\alpha \approx 1/4$).  
We observe the well known order-disorder transition in the VM
with both boundary conditions, but the application of a desired direction
condition inhibits the existence of disorder as expected. Similar behavior is
observed in the SFM case. For the SFM+VM case we find a susceptibility maximum
which increases with system size as a function of noise strength indicative of
a order-disorder transition in the whole range of densities ($\rho \epsilon [\frac{1}{12},\frac{1}{9}]$) and speeds ($v_{0} \epsilon [0.5, 2]$) studied. 

From our results we conclude that the new SFM+VM model is a well-suited model 
to describe non-panic evacuation with diverse degrees of disorder.

\end{abstract}

\maketitle

\section{Introduction}
\label{sec:intro}

Collective behavior of a large number of self-propelled particles (SPP) can
result in spontaneously developping ordered motion governed by changes in some
control parameter(s).
Many groups of living beings (from cells and bacteria to fish, birds, mammals
and even humans) exhibit this specific kind of motion under particular
conditions.  
From the point of view of statistical physics, the occurrence of collective motion is a
non-equilibrium phase transition which has attracted
much attention of the community in the last decades (for more details see the
reviews \cite{Castellano,reviewVicsek,ginelli}).  

The first step in the understanding of this complex behavior has been to
propound simple but non-trivial models, such as the Vicsek Model (VM)
\cite{VM} in the middle of the nineties. In the VM, point particles with
constant speed interact with each-other only by trying to align their direction of
motion with their nearest-neighbors, with an uncertainty of this process represented by an external
noise $\eta$. This simple rule (explained in detail in section \ref{sec:vm})
is enough to reproduce flocking behavior at low values of $\eta$, as
commonly observed in nature. 

However, the complexity of collective motion often requires to go beyond a
point particle model by taking into account short- and long- range
particle-particle interactions and interactions with the environment.  
In particular, it is possible to model a crowd as a system of particles by
considering person-person and person-wall interactions. 
One of the first models of this type proposed to describe pedestrian
motion was the so-called \textit{Social Force Model} (SFM) \cite{SFM}. 
Unlike the simplistic Vicsek Model, the SFM introduces the idea of social
interactions by modeling the individual reaction to the effect of environment
(either other pedestrians or borders), and a preferential direction of
motion. This model will be explained in detail in Sec. \ref{sec:sfm}. 

The main idea of our work is to analyze $-$by using statistical mechanical
tools$-$ the evacuation of people along a hallway in a non-panic regime, such
as occurs in schools, hospitals or airports.  
To model this situation, it is important to take into account a series of
considerations such as: the excluded-volume and mass of human-particles; the
interpersonal interactions that make them want to keep their own space; the
intent to remain away from the walls; the existence of a desired direction of
motion; but also the idea of being influenced by the motion of
nearest-neighbors. Keeping this in mind, we propose in the present work a new
approach by introducing a model $-$a combination of the standard VM and SFM$-$,
which takes into account the particle-interactions as in the SFM and additionally a
Vicsek-like alignment modulated by a noise $\eta$. We call this the
\emph{SFM+VM} model, and it will be described in detail in
Sec. \ref{sec:comb}. 

In order to identify the influence of the different parts of the model, we
have compared the VM, the SFM+VM, and the SFM in 
the stationary configuration of $N$ particles moving through a corridor-like system
in a normal evacuation situation (slow speed regime). 
We have studied all models under the same external
conditions (number of particles $N$, system-size $L_x \times L_y$, particle
speed $v_0$, external noise $\eta$, etc.). 
In particular, we have analyzed the VM under different boundary conditions,
with the purpose of introducing the effect of walls into the VM, and the idea
of a desired direction of motion, such as it is defined in the SFM, in order
to flesh out the comparison between models, and we have studied the effect of
these variations on the order-disorder phase transition.  
Finally, it is worth mentioning that even though the SFM has been widely studied
(see for example  \cite{Helbing2011,Chen} and references therein), the
statistical physics issues related to phase transitions of social models have been less
extensively studied (see e.g. \cite{Cambui,baglietto11}).

The paper is organized as follows: after this introduction, a detailed 
description of the models can be found in section \ref{sec:mod}, the
simulation details are presented in section \ref{sec:sim}, and results are
analyzed and discussed in section \ref{sec:res}. Finally, a summary and our
conclusions are developed in section \ref{sec:conc}.  

\section{Models}
\label{sec:mod}
\subsection{Vicsek Model (VM)}
\label{sec:vm}

The Vicsek Model \cite{VM} describes the dynamics of $N$ SPP characterized at
time $t$ by their position $\mathbf{r}_i(t)$ and velocity $\mathbf{v}_i(t)$
($i = 1,...,N$), and in its simplest version all particles are considered to
have the same speed $v_0$ ($\left| \mathbf{v}_i \right| =v_0$).  
At each time step, particle $i$ assumes the average direction of motion of its
neighbors (within an interaction circle of radius $R_0$) distorted by the
existence of an external noise of amplitude $\eta$ ($\eta=[0,1]$). 
The simple update rules in the 2D case are given by
\begin{align}
\theta_i(t+\Delta t)&= \langle \theta(t)\rangle_{R_0} + \eta \xi_i(t), \label{eq:thetaVM}\\
\mathbf{v}_i(t+\Delta t)&=v_0 (\cos \theta_i(t+\Delta t), \sin \theta_i(t+\Delta t)), \label{eq:velVM}\\
\mathbf{r}_i(t+\Delta t)&= \mathbf{r}_i(t) + \Delta t \; \mathbf{v}_i(t+\Delta t), \label{eq:rFU}
\end{align}
where $\langle \theta(t)\rangle_{R_0}$ is the average of the direction of
motion of all the nearest-neighbors of the $i-$th particle (whose distance
$\Vert r_j-r_i\Vert\leq R_0$), and $\xi_i(t)$ is a scalar noise uniformly
distributed in the range $[-\pi,\pi]$. The update rules given by
Eqs. \ref{eq:thetaVM} and \ref{eq:rFU} are known in the literature as
\emph{angular noise} \cite{VM} and  \emph{forward update} \cite{chate},
respectively. 
By choosing  $\Delta t=1$ as the time unit and  $R_0 = 1$ as the length
unit, the only control parameters of the model are the noise amplitude $\eta$,
the speed $v_0$ and the density of particles $\rho = N/V$, where $V=L_x\times
L_y$ is the volume of the 2D system. 

These simple rules are enough to reproduce a fundamental aspect of collective
behavior: the existence of a phase transition between a disordered state and
an ordered phase (where the direction of motion is the same for all particles)
as the noise intensity $\eta$ decreases.  
The average velocity of the system, defined as
\begin{equation}
\varphi \equiv \frac{1}{N v_0} \mid \sum_{i=1}^{N} \mathbf{v}_i \mid,
\label{opVM}
\end{equation}
is the appropriate order parameter to describe this phase transition
\cite{VM,reviewVicsek}. In the disordered phase $\varphi \sim 0$ while
$\varphi \rightarrow 1$ for the ordered phase.
A critical line $\eta_c(v_0,\rho)$ separates the disordered from the ordered states.
In order to ensure the independence from initial conditions, $\varphi$ has to be
averaged over time after it reaches a stationary regime. This value is called
$\varphi_{stat}$. The order of this phase transition is still a matter of controversy
\cite{chate,baglietto2}.  Until now the consensus seems to indicate that
depending on how the noise is applied $-$angular noise or vectorial noise$-$
the transition can be continuous or first-order like, respectively (for more
details see \cite{reviewVicsek}).

\subsection{The Social Force Model}
\label{sec:sfm}
Contemporary to the VM, the \emph{Social Force Model} (SFM) was proposed by
Helbing and Molnar \cite{Helbing91,SFM} to describe the behavior of
pedestrians. Since then, the SFM has been widely studied and applied for
different situations (see for example \cite{SFM-FULL,Helbing2001,Helbing2011}).  

The SFM determines the direction of motion for each particle by taking
into account three interactions: the "Desire Force", the "Social Force", and
the "Granular Force", which are defined as  
\begin{enumerate}
\item The "Desire Force" ($\mathbf{F_{Di}}$) represents the desire of SPP to
  march in a given direction; if we are modeling the evacuation of a
  crowd, the target of the desire will be the exit. This force involves the
  idea of a desired speed of motion $v_{D}$, and it is given by  
\begin{equation}
\mathbf{F_{Di}}=m_{i} \frac{(v_{D} \mathbf{\hat{e_{t}}} - \mathbf{v_{i}})}{\tau_{RT}}, 
\label{eq:DF}
\end{equation} 
where $m_{i}$ is the particle mass, $\mathbf{v_{i}}$ is the current velocity,
$\mathbf{\hat{e_{t}}}$ the unit vector pointing to the target direction and
$\tau_{RT}$ is the relaxation time of the particle velocity towards $v_{D}$. 
In the present work, the desired speed has been taken as equal to the
Vicsek-particle speed $v_D=v_0$.

\item The "Social Force" ($\mathbf{F_{Si}}$) takes into account the fact that
  people like to move without bodily contact with other individuals. The "private
  space" wish is represented as a long range repulsive force based on the
  distance $r_{ij}=\Vert \mathbf{r_{i}}-\mathbf{r_{j}}\Vert$ between the
  center of mass of the individual $i$ and its neighbor $j$. The complete
  expression is the following: 
\begin{equation}
\mathbf{F_{Si}} =\sum_{j(\neq i)}^{N} A \exp \left[\frac{(d-r_{ij})}{B} \right] \mathbf{\hat{n}_{ij}},
\label{eq:SF}
\end{equation}      
where the constants $A_{i}$ and $Ḅ_{i}$ define the strength and range of the social
force, $\mathbf{\hat{n}_{ij}}$ is the normalized vector pointing from
pedestrian $j$ to $i$, and $d$ is the pedestrian diameter when one considers
identical particle sizes.  

\item The "Granular Force" ($\mathbf{F_{Gi}}$) is considered when the
  pedestrians are in contact each other. It is a repulsive force inspired by
  granular interactions, includes compression and friction terms, and is
  expressed as 
\begin{equation}
\mathbf{F_{Gi}} =  \left[ k \hspace{1mm} \mathbf{\hat{n}_{ij}} + \kappa
  \Delta v^{t}_{ji} \hspace{1mm} \mathbf{\hat{t}_{ij}}\right] 
g(d-r_{ij}). 
\label{eq:GF}
\end{equation}   
where $g(d-r_{ij})$ is zero when $d<r_{ij}$ and $d-r_{ij}$
otherwise. The first term represents a compressive force, its strength given by the
constant $k$, which acts in the $\mathbf{\hat{n}_{ij}}$ direction.    
The second term in Eq. \ref{eq:GF} $-$related to friction$-$ acts in the
tangential direction $\mathbf{\hat{t}_{ij}}$ (orthogonal to
$\mathbf{\hat{n}_{ij}}$ ),
 and it depends on the difference $\Delta
v^{t}_{ji}=(\mathbf{v_{j}}-\mathbf{v_{i}}) \cdot \mathbf{\hat{t}_{ij}} $
multiplied by the constant $\kappa$.  
\end{enumerate}

The interaction pedestrian$-$wall is defined analogously by means of social
($\mathbf{F_{SWi}}$) and granular forces ($\mathbf{F_{GWi}}$).  
If $r_{iW}$ denotes the distance between the $i$-pedestrian and the wall, and
$\mathbf{\hat{n}_{iW}}$ is the wall normal pointing to
the particle, the ''Social force'' is defined as 
 \begin{equation}
\mathbf{F_{SiW}} = A \exp \left[\frac{(d/2 - r_{iW})}{B} \right] \mathbf{\hat{n}_{iW}}.
\label{eq:SFW}
\end{equation}

Similarly, denominating $\mathbf{\hat{t}_{iW}}$ as the direction tangential
(orthogonal to $\mathbf{\hat{n}_{iW}}$), the "Granular force" is expressed as 
 \begin{equation}
\mathbf{F_{GWi}} =  \left[ k \hspace{1mm} \mathbf{\hat{n}_{iW}} - \kappa
  (\mathbf{\hat{v}_{i}} \cdot \mathbf{\hat{t}_{iW}}) \hspace{1mm} \mathbf{\hat{t}_{iW}}\right] g(r-r_{iW}). 
\label{eq:GFW}   
\end{equation}
For the force constants in the interactions between the particles and the
walls we choose the same values as for the interparticle forces.

By considering all the forces described above, the equation of motion for 
pedestrian $i$ of mass $m_i$ is given by  
\begin{equation}
m_i \frac{d \mathbf{v_{i}}}{dt} = \mathbf{F_{Di}} + \mathbf{F_{Si}} + \mathbf{F_{SWi}} + \mathbf{F_{Gi}} + \mathbf{F_{GWi}}
\label{eq:sfm}
\end{equation}

\subsection{The combined model (SFM+VM)}
\label{sec:comb}
Realistic evacuations in a non-panic situation have different behavior
depending on geometry and average speed. However, and even in the case of similar
boundary conditions, it is expected that one observes differences between
evacuations in a school, hospital or airport. An important factor here is the
existence of intrinsic fluctuations in the moving-interacting particles, such
as children, passengers or patients.  
We propose to introduce these fluctuations including in the standard SFM the
external noise parameter $\eta$ as in the Vicsek model. 

The central idea is to take into account not only the social interactions
described above but the influence of nearest-neighbors in the
\emph{Vicsek-style}, and to include noise $\eta$ as an external parameter,
which modulates this interaction. 
We will refer to this new model as \emph{SFM+VM}.
In this way, in the SFM+VM the velocity of particle $i$ is given by
\begin{equation}
\mathbf{v}_{i}(t+\Delta t) = v_0 \frac{\mathbf{v}_{VM_{i}}(t+\Delta t) +  \frac{d
    \mathbf{v}_{i}}{dt}(t) \Delta t}{\Vert \mathbf{v}_{VM_{i}}(t+\Delta t) +
  \frac{d\mathbf{v}_{i}}{dt}(t) \Delta t\Vert}.
\label{eq:sfmvm}
\end{equation}
Here $\mathbf{v}_{VM_{i}}(t+\Delta t)$ is the 
velocity of particle $i$ given by Eq.(\ref{eq:velVM}).

\section{Simulation Details}
\label{sec:sim}
As was previously mentioned, we are going to compare all models  under equal external conditions.  
An important point here is to match all relevant physical units in the models.
Since $R_0$ and $\Delta t$ are common variables, we can extend their role as
time and length units from the VM to all models studied. 
In order to match the 
simulations to real systems, we have assumed that one length-unit ($R_0$)
is equivalent to one meter (in SI), and one second corresponds to one time step
$\Delta t$. After this assumption, it is possible to
define the mass and force variables in Eq.(\ref{eq:sfm}) as kilograms and
Newtons, respectively, as the SFM requires.   

Monte Carlo simulations were performed in a system of $N= 300$ self-propelled
particles moving in corridor of size  $L_x \times L_y$, with $L_{x} = 600$ and
$L_{y}\in [2.5,6]$.    
Periodic boundary conditions were applied in the horizontal direction at
$L_{x}$, so that circulation of particles occurred in a loop. 
A size of $L_{x} = 600$ was then chosen to make sure that the fastest
particles do not meet the stragglers.

To correlate pedestrians with particles, in the case of SFM+VM, we have
considered particles with $d=0.7$ (in units of $R_0$) and 80~kg of mass.     
The characteristic parameters of SFM interactions
(Eqs. \ref{eq:DF}--\ref{eq:GF}) have been taken from previous works
\cite{SFM,Frank-Dorso}, specifically  $A$ = 2000~N, $B$ = 0.08~m, $k=
1.2 \times 10^{5}$ kg s$^{-2}$ and $\kappa = 2.4\times 10^{5}$ kg/(m s). These
values correspond to a typical crowd.

In every case studied, several tests have been performed to assure the
reliability of the data, that are not shown here for the sake of space.  
We made sure that starting from different initial conditions of particle distribution in
position and direction of motion, $\varphi (t)$ reached the same value in the
stationary regime ($\varphi_{stat}$).  
Also, the run-to-run fluctuations in the $\varphi (t)$ profile were not
drastic. To determine the number of reasonable runs for the simulations, a
first study was made observing how the average value of $\varphi_{stat}$ and
its uncertainty varied according to how many runs were taken in the
average. As a conclusion, we considered $50$ runs for each set of simulation parameters.  

As was mentioned, the VM is a simple model that does not take into account
short-range interactions such as excluded-volume or friction between particles
and with walls. Moreover, taking into account that the aim of this work is to
analyze the collective motion of individuals moving through a corridor, we
introduce a series of variations on the VM. 
On the one hand, we relaxed the standard periodic boundary conditions (PBC)
to a bouncing-back condition (BbC) in the y-direction.  
In this way, we simulate impenetrable walls at $y=0$ and $y=L_y$, where
particles rebound without losing energy.  

On the other hand, we have introduced a \textit{desired-direction} (DD) of
motion, in such away that the direction of motion at time $t+1$ given by
Eq.(\ref{eq:thetaVM}) is modified by the addition of a desired-angle
$\theta_{des}$ as 
\begin{equation}
\tilde{\theta_i}(t+\Delta t)= \frac{\theta_i(t+\Delta t)-\theta_{des}}{2},
\label{eq:angdes}
\end{equation}
with $\theta_{des}=0$ in our case, indicating that particles prefer to move to the end of the corridor.
This variation of the VM allows to introduce the existence of a preferred
direction of motion, in the same sense as it is introduced in the SFM.

Finally, particles were inserted randomly in the first 0.5 $L_x$ of the
corridor with the intention to move towards the end of the corridor. The speed
was considered in the range $v_0=[0.5,2]$ (m/s), consistent with normal
evacuations in schools, hospitals, cinemas, etc. 

For this reason, we have explored the range of $L_{y}\in [2.5,6]$ (m) in order to
represent the typical widths of corridors.

\section{Results and discussion}
\label{sec:res}

\begin{figure}[!h]
\centering
\includegraphics[width=\columnwidth]{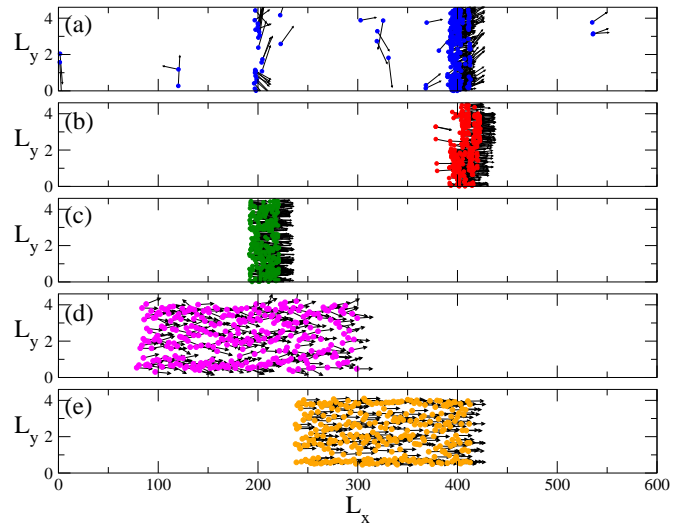}
\caption{Snapshots in the stationary-ordered state for $N=300$, $L_x=600$,
  $L_y=4.5$ ($\rho=1/9$), $\eta=0.05$, $v_0=0.5$, and for different cases
  studied: $(a)$ VM with periodic boundary conditions in the $y$-direction (PBC);
  $(b)$ VM with bouncing-back boundary condition in the $y$-direction (BbBC);  $(c)$ VM+BbBc with
  a desired direction of motion (DD); $(d)$ SFM+VM; and $(e)$ SFM (in this
  case, the external noise $\eta$ is not defined in the model). In all cases,
  PBC were applied in the $x-$direction. (Color on-line: different colors link
  with data of Fig. \ref{fig4})} 
\label{fig1}
\end{figure}

A full set of simulations was performed by taking into account the
considerations mentioned above. In Figure \ref{fig1} we show several
snapshots of stationary configurations in the ordered phase 
for all models with the same external parameters ($\eta=0.05$ and $v_0=0.5$)
and different boundary conditions.

As can be seen, in the VM (Figure \ref{fig1} (a--c)),
stationary-ordered-states correspond to the existence of bands perpendicular
to the direction of motion. This fact has been widely studied and reported in the
literature (see \cite{reviewVicsek,ginelli}, and reference therein). 
However, it is worth to mention that different boundary conditions seem to
affect the local order within the band of particles (e.g., PBC in Figure
\ref{fig1}(a) in comparison with BbBC in Figure \ref{fig1}(b)).  
Following with the qualitative analysis of the VM cases, the most ordered
configuration corresponds to the VM+BbBC+DD, as it is expected.  
Here, the incorporation of a preferential direction of motion can be
interpreted as an external field applied in the system promoting order.
On the other hand, in both SFM+VM at low noise (Figure \ref{fig1}(d)) and SFM (Figure \ref{fig1}(e)), particles are ordered in a horizontal cluster configuration trying to keep a distance between each other and with the walls.

\begin{figure}[!h]
\centering
\includegraphics[width=\columnwidth]{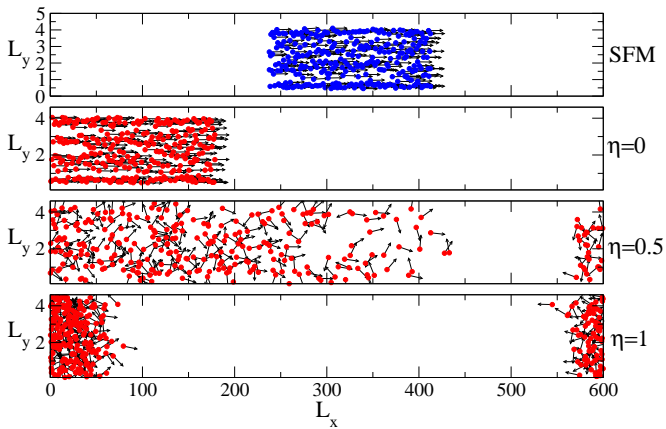}
\caption{Comparison of the stationary-state configurations for both SFM and
  SFM+VM for different external noise values $\eta$, as indicated. The snapshots
  correspond to $N=300$, $L_x=600$, $L_y=4.5$ ($\rho=1/9$), and $v_0=0.5$.} 
\label{fig2}
\end{figure}

The similarity in behavior observed in snapshot configurations between low noise
SFM+VM (Figure \ref{fig1}(d)) and SFM (Figure \ref{fig1}(e)) is
broken when the noise $\eta$ increases.  
This is explicitly observed in Figure \ref{fig2}, where, as it is
expected, disorder increases with noise $\eta$. 

To appreciate in detail the influence of Vicsek interactions on the spatial
configuration in the SFM, we have studied the evolution of particle clusters for several
noise strengths.   
For this purpose, we define the density profile in the x-direction as 
\begin{equation}
\text{P} (x,t) = \frac{\text{Number of particles between $x$ and $x+\Delta x$}}{N},
\end{equation}
and consequently $\text{P} (x,t)$ can be interpreted as a histogram, with bins of width $\Delta x$.
In our case, we have fixed $\Delta x=5\gg d=0.7$ so that we have the chance to
find many particles in the bin. The size of the bin is such that it is not too
small to allow the cluster to be described as continuous (no empty bins in
the middle) but also not so large that it cannot be appreciated how the size
changes over time.  
 
In this way, the width of the density profile at time $t$ ($w(t)$) can be defined as the distance between the maximum and the minimum values of $x$ for which $P(x,t)> 1/N$, properly normalized by considering the PBC applied in the $x$-direction.
With this idea, $w(t)$ is directly associated with the extension of the cluster of particles as a function of time.

\begin{figure}[!h]
\centering
\includegraphics[width=\columnwidth]{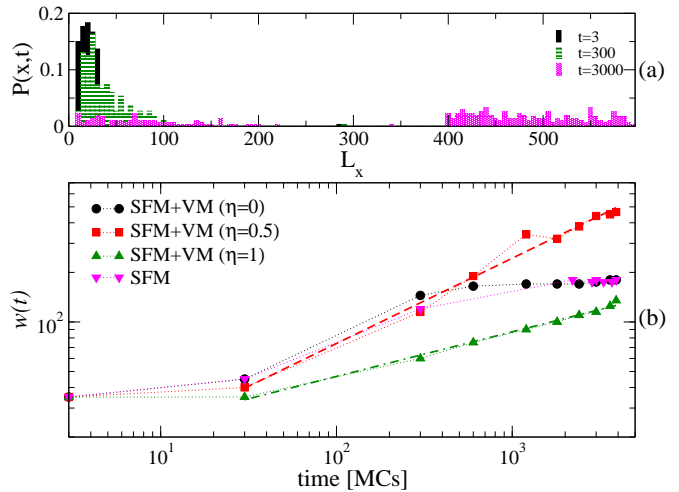}
\caption{$(a)$ Density profile of the SFM+VM for $\eta=0.5$, $N=300$, $L_x=600$,
  $L_y=4.5$ ($\rho=1/9$), and $v_0=0.5$. $(b)$ Log-Log plot of the cluster width $w(t)$
  versus time for $N=300$, $L_x=600$, $L_y=4.5$, $v_0=0.5$, and different
  noise strengths as indicated. The segmented lines represent the fits to the points
  proposing a power-law behavior.} 
\label{fig3}
\end{figure}

The obtained results show that at low noise the stationary cluster keeps its form and
$w(t)$ is constant in time. On the other
hand, when the noise
increases particles spread and therefore the cluster width grows with
time. This effect is most relevant for $\eta=0.5$. 
The dynamic dependence of $w(t)$ (Figure \ref{fig3} (b)) suggest a power-law behavior of the form 
\begin{equation}
w(t) \propto t^{\alpha}, 
\end{equation}
where $\alpha$ has a strong dependence on noise.
In fact, a least-squares fit of the data gives $\alpha\approx 0$ for $\eta=0$
and the SFM, $\alpha=0.52(4)\approx \frac{1}{2}$ for $\eta=0.5$, and
$\alpha=0.27(1)$ for $\eta=1$. The behavior observed for $\eta = 0.5$ is
compatible with a diffusive-like spread of the particle front with the typical
Einstein exponent $\alpha = \frac{1}{2}$. For higher noise it seems
that the competition between random movement (given by the noise) and
social-force interactions gives as a result a sub-diffusive behavior with
$\alpha \approx \frac{1}{4}$. 
The behavior observed for $\eta=1$ is reminiscent of the \emph{freezing by
  heating} effect observed in the SFM in the panic-regime
\cite{PhysRevLett.84.1240,Helbing2002SimulationOP}.  
In this case, the existence of increasing fluctuations in the system, or
\emph{nervousness of pedestrians}, produces a blocking effect and
even when they are in a disordered state particles can not move in the desired direction
of motion.

Let us now turn to the question of an underlying non-equilibrium phase
transition in the different models as a function of the noise strength by
evaluating the stationary state $\varphi_{stat}$ and its 
variance $\text{Var}(\varphi)\equiv \langle \varphi^2\rangle -
\langle\varphi\rangle^2$.

\begin{figure}[!h]
\centering
\includegraphics[width=\columnwidth]{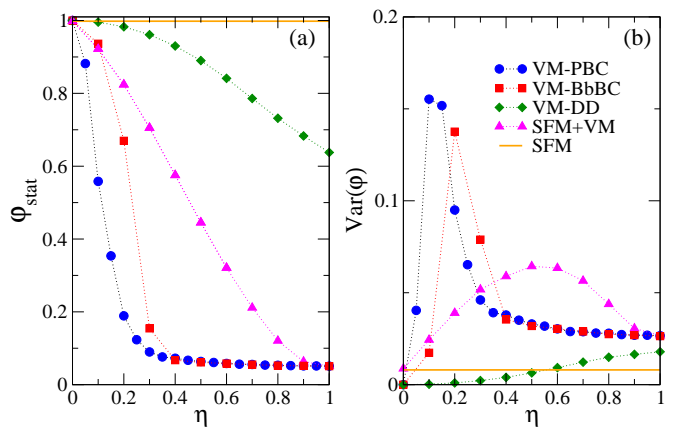}
\caption{$(a)$ Order Parameter $\varphi_{stat}$, and $(b)$ variance
  $\text{Var}(\varphi)$, as a function of external noise $\eta$, for $N=300$,
  $L_x=600$, $L_y=4.5$, $v_0=0.5$, and different models studied, as indicated.} 
\label{fig4}
\end{figure}

The dependence of both quantities ($\varphi_{stat}$ and $\text{Var}(\varphi)$)
as a function of $\eta$ for fixed speed ($v_0=0.5$) and lattice size
($L_x\times L_y=600\times 4.5$) can be observed in Figure \ref{fig4}. 
As a first comment, it should be noticed that the application of BbBC seems to
move the VM transition (maximum $\text{Var}(\varphi)$ in
Fig. \ref{fig4}(b)) to higher values of $\eta$ in comparison with PBC.  
This is in agreement with the snapshots of Figure \ref{fig1} (a-b); at a
given noise ($\eta<\eta_c$) the PBC case is more disordered than the BbBC one.  
Even when in both cases the stripe geometry confines particle movement, the
existence of impenetrable walls in the BbBC increases significantly the confinement
effects, and therefore it is expected that VM-alignment should be
more relevant than for PBC.

The application of a desired direction of motion (DD) in the VM has substantially
different consequences.  Unlike in the standard VM (with both PBC and BbBC), VM+DD
prevents the existence of a disordered phase as it is expected. This appears
reflected in the fact that $\varphi_{stat}\nrightarrow 0$, and
$\text{Var}(\varphi)$ is maximal when $\eta \rightarrow 1$. 

In the case of the SFM+VM the presence of an external noise clearly modifies the
behavior of motion in comparison with SFM in a non-panic regime in a corridor
(when trivially one expects $\varphi_{stat}=1$).  
Because in the SFM+VM repulsive interactions between particles make the
formation of a condensed cluster more difficult, it is observed that $\varphi_{stat}^{VM+DD}>
\varphi_{stat}^{SFM+VM}$ for every $\eta$. The existence of a maximum in $\text{Var}(\varphi)$
(Fig. \ref{fig4}(b)) suggests to
analyze the dependence of this behavior on external noise upon a variation of speed
$v_0$ and system size $L_x \times L_y$.  
To check the reliability of the SFM+VM outcomes, we have performed the same
analysis in VM cases, where the $\text{Var}(\varphi)$ maximum is related to
the existence of a phase transition.

\begin{figure}[!h]
\centering
\includegraphics[width=\columnwidth]{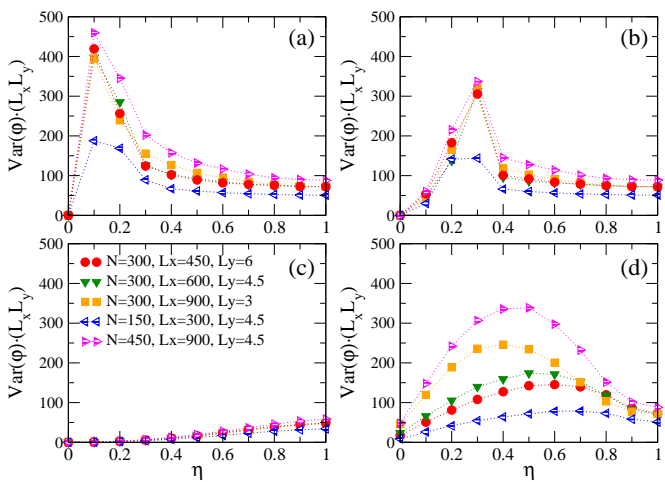}
\caption{$\text{Var}(\varphi) \cdot (L_x L_y)$ as a function of noise $\eta$
  for fixed density $\rho=1/9$, $v_0=0.5$, and different system size
  $L_x\times L_y$ and number of particles $N$, as indicated for $(a)$ VM with
  PBC, $(b)$ VM with BbBC, $(c)$ VM with BbBC and desired direction of motion
  (DD), and $(d)$ SFM+VM.} 
\label{fig5}
\end{figure}

Our results  are shown in Fig. \ref{fig5}. As can be seen, in the VM
(Fig. \ref{fig5}(a-b)) the peak of the susceptibility $-$defined as
$\text{Var}(\varphi) \cdot (L_x L_y)$ $-$ becomes narrower and higher as both
$N$ and system size are increased.  
This behavior is not observed in the case of  VM+DD (Fig. \ref{fig5}(c)), as expected.
In the case of the SFM+VM the --rounded-- peak of $\text{Var}(\varphi) \cdot
(L_x L_y)$ as a function of $\eta$ increases in height as $N$ and the system
size grow, however, it sharpens only very slowly. This is indicative of an
underlying phase transition but to clearly establish its
existence it would be necessary to study much larger system sizes at fixed
density than is possible within the scope of the present work.   
However, it is noteworthy that although the effect of external noise
$\eta$ in the model is less important than in the VM, it seems enough to break the
symmetry imposed by SFM interactions.

\begin{figure}[!h]
\centering
\includegraphics[width=\columnwidth]{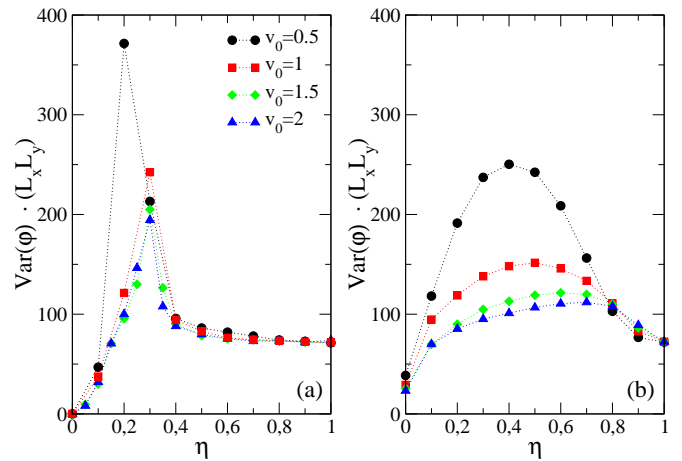}
\caption{$\text{Var}(\varphi) \cdot (L_x L_y)$ as a function of noise $\eta$
  for $N=300$, $L_x=600$, $L_y=4.5$ and different speeds $v_0$ as
  indicated. Cases are VM with BbBC $(a)$, and SFM+VM $(b)$.} 
\label{fig6}
\end{figure}

Similar behavior is observed for the dependence of  $\text{Var}(\varphi) \cdot
(L_x L_y)$ on speed $v_0$ (see Fig. \ref{fig6}). 
The $\text{Var}(\varphi)$ maximum diminishes as the speed increases, and its
position seems to move to $\eta = 1$ for larger $v_0$. 
As can also be seen in this figure, for a given speed $v_{0}$ the
maximum values of  $\text{Var}(\varphi) \cdot (L_x L_y)$ ​​are higher in the case of
the VM+BbBC than the SFM+VM.

\begin{figure}[!h]
\centering
\includegraphics[width=\columnwidth]{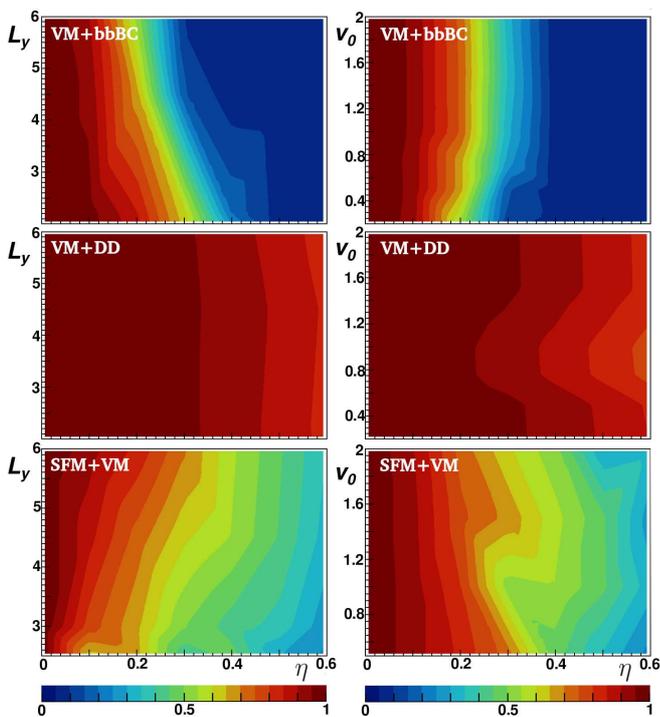}
\caption{(Color-online) Level plots of $\varphi_{stat}$ for $N=300$, $L_x=600$
  as a function of noise $\eta$ (x-axis). 
\emph{Left-pannel:} for different densities ($L_y$ variable and $L_x$ fixed)
for fixed speed $v_0=0.5$. \emph{Right-pannel:} for different speed $v_0$, and
at fixed $L_y=4.5$ ($\rho=1/9$).} 
\label{fig7}
\end{figure}

Finally, substantial information can be gleaned from level plots of
$\varphi_{stat}$ for several $L_{y}$ values and noise values in the range
$0-0.6$ (Fig. \ref{fig7}(a-c)). Because in these plots we are keeping
$N$ and $L_{x}$ fixed, the vertical axis $L_{y}$ is an indirect representation
of the density $\rho$.  
From these plots it can be appreciated that although the previous analysis was
presented for a given value of $\rho$ and $v_{0}$, similar behavior is observed for
$\rho \epsilon [\frac{1}{12}, \frac{1}{5}]$ and $v_{0} \epsilon [0.5, 2]$
(Fig. \ref{fig7}).  
In this way, our conclusions can be extended to a wide range of densities and
speeds within the non-panic regime.

\section{Summary and Conclusions}
\label{sec:conc}

We have studied the role of interactions in the behavior of pedestrians moving
in a corridor-like system. For this purpose, we have introduced a new model
that we have called SFM+VM, as a combination of both the well-known Vicsek and
Social Force models. To check its performance, we have started our analysis
with the Vicsek model in a confined geometry with different boundary conditions
applied. In particular, we have studied the effects of bouncing-back boundary
conditions that reproduce the effects of walls, and the existence of a
desired direction of motion -the end of the corridor. We have compared these
results to those obtained with the SFM, and the new SFM+VM. 

In the first place, particle configurations in the ordered-steady-state are
qualitatively different between the models analyzed. 
While in the VM, particles move in a more or less compact band perpendicular
to the direction of motion, in the SFM and the SFM+VM particles exhibit some
horizontal stripes parallel to the direction of motion. This effect is a
consequence of the repulsive interactions between particles and with the walls
present in those two models (Figures \ref{fig1} and \ref{fig2}).  
In order to compare both SFM and SFM+VM, we have analyzed the density-profiles
in the direction of motion ($x$) and determined its width as a function of
time. Our results indicate that the width in the SFM+VM case has a power-law
behavior with a dynamical exponent $\alpha$, which depends on the external
noise $\eta$. In particular, we have determined that $\alpha \approx 1/2$ at
$\eta=0.5$ and  $\alpha \approx 1/4$ at $\eta=1$ (Figure \ref{fig3}). We
have associated this change from an expected diffusive-like to a sub-diffusive
behavior to the competition between VM-like interactions and social
interactions.  
At high noise, fluctuations have a \emph{freezing by heating} effect, that has
been reported only in the panic-regime before
\cite{PhysRevLett.84.1240,Helbing2002SimulationOP}.  
In our case, this effect appears in the system even at low-speed values, as a
consequence of the introduction of the external noise.

In the second place, we have analyzed the order parameter $\varphi$, defined
as the average velocity of the system, and its variance as function of
external noise $\eta$ for the diverse models described above (Figures
\ref{fig4}$-$\ref{fig5}). In the VM, the application of different
boundary conditions (periodic and bouncing-back) moves the critical value of
the order-disorder phase transition to higher values of $\eta$. In contrast, the
existence of a desired direction of motion in the VM  promotes the order even
at high values of external noise, annihilating the phase transition as
expected. In the SFM+VM, the existence of an external noise that modulates the
Vicsek-like interactions brakes the SFM symmetry (Figure \ref{fig6}). As
a consequence, both the order parameter and its variance are sensitive to this
effect.  Finally, we consider it important to remark that these outcomes can be
observed in the whole range of densities $\rho \hspace{1mm}
\epsilon \hspace{1mm} [\frac{1}{12},\frac{1}{9}]$ and speeds
$v_{0} \hspace{1mm} \epsilon \hspace{1mm} [0.5, 2]$ studied, which encompass reasonable values of
evacuations in a non-panic regime (Figure \ref{fig7}).  

Based on these results, we can conclude that the SFM+VM is a successful model to
describe the pedestrian motion along a corridor in a non-panic regime.  
This new model allows us to elucidate the role of the competition between
social and alignment interactions, characteristics of  the SFM and the VM,
respectively.  
In the SFM+VM, alignment interactions are tuned by the external noise $\eta$
same as for the VM and this allows us to
address questions on the existence of a non-equilibrium order-disorder
transition controlled by this parameter. Our results are qualitatively
compatible with the existence of such a transition, however, the approach to
the thermodynamic limit seems to be very slow putting a final quantitative
conclusion beyond the scope of this work.

\acknowledgements

This work was supported by Consejo Nacional de Investigaciones
Cient\'\i{}ficas y T\'ecnicas (CONICET), Universidad Nacional de La Plata
(Argentina), and Universidad Nacional de Quilmes (Argentina). Simulations were
done on the cluster of \emph{Unidad de C\'alculo}, IFLYSIB. 
We also thank Martin-Luther-University Halle-Wittenberg and Alexander von
Humboldt Foundation for financial support.

\bibliographystyle{aipnum4-1}
\bibliography{bibliog}

\end{document}